\documentclass[aip,apl,reprint]{revtex4-1}
\draft
\usepackage{graphicx}
\usepackage{subcaption}
\usepackage{amsmath}

\usepackage{fancyhdr}
\fancypagestyle{plain}{%
 \fancyhf{} 
 \fancyfoot[C]{DISTRIBUTION STATEMENT A. Approved for public release: distribution unlimited.}
 
}
\fancypagestyle{firstpage}{%
\fancyhf{}\lfoot{DISTRIBUTION STATEMENT A. Approved for public release: distribution unlimited. AFRL-2024-0867}}

\begin{document}
\title{Infrared Vertical External Cavity Surface Emitting Laser Threshold Magnetometer}
\author{Nathan~S.~Gottesman}
\email{nathangottesman@arizona.edu}
\affiliation{Wyant College of Optical Sciences, University of Arizona, 1630 E. University Blvd., Tucson, Arizona 85721, USA}
\affiliation{DeUVe Photonics, 2030 N. Forbes Blvd., Tucson, Arizona 85745, USA}
\author{Michael~A.~Slocum}
\author{Gary~A.~Sevison}
\author{Michael~Wolf}
\affiliation{Materials and Manufacturing Directorate, Air Force Research Laboratory, Wright Patterson Air Force Base, Ohio 45433, USA}
\author{Michal~L.~Lukowski}
\author{Chris~Hessenius}
\author{Mahmoud~Fallahi}
\affiliation{Wyant College of Optical Sciences, University of Arizona, 1630 E. University Blvd., Tucson, Arizona 85721, USA}
\affiliation{DeUVe Photonics, 2030 N. Forbes Blvd., Tucson, Arizona 85745, USA}
\author{Robert~G.~Bedford}
\affiliation{Materials and Manufacturing Directorate, Air Force Research Laboratory, Wright Patterson Air Force Base, Ohio 45433, USA}
\date{\today}
\begin{abstract}
Nitrogen-vacancy (NV) centers have considerable promise as high sensitivity magnetometers, however are commonly limited by inefficient collection and low contrasts. Laser threshold magnetometry (LTM) enables efficient collection and high contrasts, providing a path towards higher sensitivity magnetometry. We demonstrate an infrared LTM using an ensemble of NV centers in a single crystal diamond plate integrated into a vertical external cavity surface emitting laser. The laser was tuned to the spin dependent absorption line of the NV centers, allowing for optical readout by monitoring the laser output power. We demonstrate a magnetic sensitivity of 7.5~nT/$\sqrt{\textit{Hz}}$ in the frequency range between 10 and 50 Hz. Furthermore, the contrast and the projected PSNL sensitivity are shown to improve significantly by operating close to the lasing threshold, achieving 18.4\% and 26.6~pT/$\sqrt{\textit{Hz}}$ near threshold. What's more, an unexpected saturable absorption phenomenon was observed near threshold, which enhanced the contrast and projected PSNL sensitivity.
\end{abstract}
\maketitle
\thispagestyle{firstpage}

Vector magnetometry has gained interest for its applications in a variety of fields including medical monitoring and diagnostics, surveying, non-destructive integrated circuit metrology, navigation, and fundamental studies of magnetism.\cite{baillet_magnetoencephalography_2017,nabighian_historical_2005,turner_magnetic_2020,mccullian_broadband_2020,canciani_absolute_2016} Most of these applications would benefit from compact high sensitivity sensors which may operate under ambient conditions. Nitrogen-vacancy (NV) center magnetometers with large sensing volumes are expected to offer impressive broadband ($\sim$100 kHz) sensitivities with technical simplicity; however, limitations due to photon shot noise must be overcome for its potential to be realized. \cite{barry_sensitivity_2020} 
\\
The NV center is a defect in diamond composed of a substitutional nitrogen and a nearest-neighbor lattice vacancy. The NV center symmetry axis may take one of four orientations dictated by the diamond lattice. This allows for most NV sensors to be extended to vector magnetometry. NV ensembles have reached sub-pT/$\sqrt{\textit{Hz}}$ sensitivity operating under ambient conditions.\cite{fescenko_diamond_2020,gao_high_2023,silani_nuclear_2023} The photon shot noise limited sensitivity can be related to the collection efficiency, contrast, and microwave resonance linewidth. Improvements to these parameters would result in a compact, high sensitivity room temperature magnetometer.
\\
The majority of current NV magnetometry is realized by collecting the photoluminescence of the NV centers, but the large emission angle leads to low collection efficiencies. Some approaches to address this problem have been developed\cite{siyushev_monolithic_2010,le_sage_efficient_2012}, or alternatively optical magnetometry may be performed with the absorption of an infrared probe.\cite{acosta_broadband_2010,acosta_optical_2010} The directionality of a probe laser leads to collection efficiencies approaching unity; however, the small cross-section of the infrared transition requires optical enhancement cavities to increase the optical path length in the diamond.\cite{dumeige_magnetometry_2013,jensen_cavity-enhanced_2014,chatzidrosos_miniature_2017}
\\
Recently, a novel approach to NV center magnetometry called laser threshold magnetometry (LTM) has been proposed, and is expected to rival the performance of the most sensitive NV magnetometers with a clear path towards low power operation. \cite{jeske_laser_2016} LTM promises improved sensitivity due to increased contrast and high directionality resulting in significantly higher collection efficiencies.\cite{jeske_laser_2016} Several implementations of LTM have been proposed\cite{jeske_laser_2016,raman_nair_absorptive_2021,webb_laser_2021,dumeige_infrared_2019}, and magnetic field dependent light amplification by stimulated emission of NV centers in a high finesse cavity was demonstrated using a 710~nm seeding laser with a projected photon shot noise limited (PSNL) sensitivity of 29.1~pT/$\sqrt{\textit{Hz}}$.\cite{hahl_magnetic-field-dependent_2022} Alternatively, the predicted sub-pT/$\sqrt{\textit{Hz}}$ sensitivity of an infrared LTM would be comparable to the best sensitivities reported in NV center magnetometers to date.\cite{dumeige_infrared_2019}
\\
In this work, we demonstrate infrared LTM using a vertical external cavity surface emitting laser (VECSEL). Long photon lifetimes in comparison to excited state lifetimes make the VECSEL architecture an ideal candidate for LTM. This enables low noise class-A laser operation, which is characterized by suppressed relaxation oscillations in response to a change in gain or loss. This results in improved linearity between spin dependent changes in absorption and laser output power. \cite{baili_shot-noise-limited_2007,baili_direct_2009,myara_noise_2013} A single-crystal diamond plate doped with NV centers was inserted into a VECSEL lasing at 1042~nm and was illuminated with 532~nm laser light. By applying a microwave probe, magnetic field dependent laser output power was detected. Furthermore, significant improvements in both the contrast and projected PSNL sensitivity were observed when operating near threshold. We reached a projected PSNL sensitivity of 26.6~pT/$\sqrt{\textit{Hz}}$ with a contrast of 18.4\% and a magnetic sensitivity of 7.5 nT/$\sqrt{\textit{Hz}}$ limited by the VECSEL's technical noise.
\\

\begin{figure}[!t]
\centering
\begin{subfigure}{.45\linewidth}
\centering
\includegraphics[width=\linewidth]{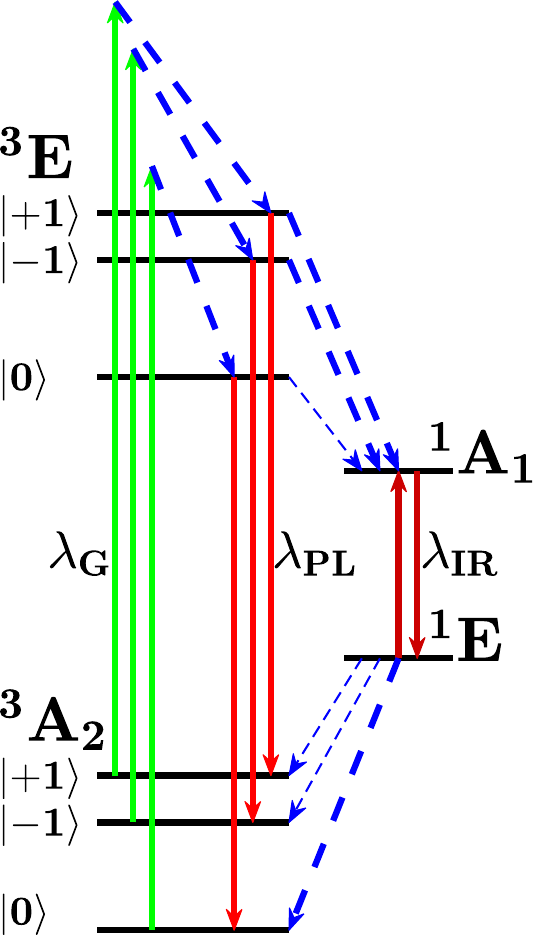}
\caption{}
\label{Energy Diagram}
\end{subfigure}%
\begin{subfigure}{.55\linewidth}
 \centering
 \includegraphics[width=\linewidth]{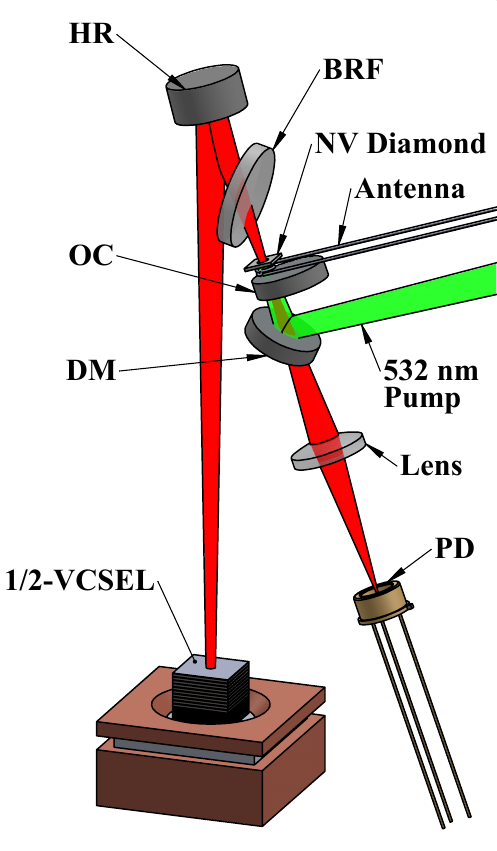}
 \caption{}
 \label{schematic}
 \end{subfigure}
 \caption{(a) Schematic of the relevant energy levels and transitions for infrared absorption based magnetometry. The transition between two spin triplet states ($^3A_2$ and $^3E$) separated by a zero-phonon line of 637 nm is pumped with $\lambda_G=$ 532~nm laser light. Spin-dependent intersystem crossings allow for two singlet states separated by a zero-phonon line of $\lambda_{IR}=$ 1042~nm to be populated. This results in spin-dependent absorption around 1042~nm. (b) Schematic of the VECSEL geometry. The cavity includes a $\frac{1}{2}-$VCSEL gain chip, diamond plate doped with NV centers (NV Diamond), antenna, birefringent filter (BRF), output coupler (OC), and concave high reflectivity mirror (HR). The out-coupled light is detected at the photodiode (PD). 532~nm pump light is delivered to the NV centers by the dichroic beam steering mirror (DM) which is aligned to counter propagate the 532 nm pump light through the OC.}
\end{figure}
Our approach relies on continuous wave (CW) optically detected magnetic resonance (ODMR) which may be understood by treating the NV center with a rate equation model. The NV center has a ground and excited electronic spin triplet state, $^3A_2$ and $^3E$, respectively, with the spin-conserving transition between the two having a zero-phonon line of 637 nm. Due to spin-spin interaction between unpaired electrons, the $m_s=$0 state is split from the spin $m_s=\pm1$ states by the zero-field splitting (ZFS) parameter. In the ground triplet state, the ZFS parameter is $D\approx$2.87~GHz. Under the presence of a magnetic field, the degeneracy of the $m_s=\pm1$ states is lifted by the Zeeman effect. These triplet states are coupled to two singlet states, $^1E$ and $^1A_1$, by spin-dependent nonradiative intersystem crossings. The transition energy between the singlet excited and ground states has a zero-phonon line of 1042~nm. Manipulation of the population in the ground triplet state by microwave fields may then result in magnetic field dependent absorption of 1042~nm light. The NV center energy structure is depicted in Fig.~\ref{Energy Diagram}. The diamond sample used for these experiments was a 3 mm$\times$3 mm$\times$500 $\mu$m plate grown with chemical vapor deposition by Element Six. The sample had a NV density of 300 ppb and a spin dephasing time $T_2^*=1\mu s$. The sample was antireflection coated for 1042~nm light using a single layer of Al$_2$O$_3$ to minimize effects due to surface polish and alignment of the diamond facets.
\\
The LTM presented here was a VECSEL which utilized a half-vertical cavity surface emitting laser ($\frac{1}{2}-$VCSEL) gain chip. The $\frac{1}{2}-$VCSEL was processed from a wafer grown by metal organic chemical vapor deposition on a GaAs substrate. It was composed of 12 compressively strained InGaAs quantum wells with GaAs barriers and a strain compensating GaAsP layer between each quantum well, followed by 25 alternating pairs of AlGaAs/AlAs serving as a high reflectivity ($R_s>99.9$) distributed Bragg reflector (DBR). Thicknesses were chosen to achieve resonant periodic gain. The DBR was soldered to a poly-crystalline diamond plate also produced by Element Six to act as a heat spreader. The GaAs substrate was removed by chemical wet etching, exposing an optically flat air-semiconductor interface. The chip was anti-reflection coated with a single layer of SiO$_2$ by electron beam evaporation for the 808~nm pump laser light. The chip was then clamped to a copper cold plate, which was maintained at 19.75 $^\circ$C with a Peltier cooler. The external cavity of the VECSEL was composed of the diamond sample previously described, a 1 mm thick birefringent filter (BRF), a concave (50 mm radius of curvature) high reflectivity mirror (HR), and a plane output coupling mirror (OC) in a 'V-shaped' cavity geometry. This cavity geometry was chosen to achieve a tight focus ($2w_s\approx 80 \mu m$) of the TEM$_{00}$ mode in the diamond sample to limit the number of surface imperfections interacting with the lasing mode, and reduce the required power of the 532~nm pump laser light used to spin polarize the NV centers. The OC was coated to be partially reflective for 1042~nm light ($R_s=97\pm 0.5\%$) and low reflectivity for 532~nm light ($R_g<3\%$). The 532~nm pump laser was coupled into the NV centers by counter propagation through the OC. This was accomplished by focusing the pump light using a plano-convex 100 mm focal length lens, then directing the 532 nm pump light through the output coupler with a dichroic steering mirror to the NV-doped diamond sample. The laser output of the VECSEL passed through the dichroic steering mirror and was focussed onto a silicon avalanche photodiode (APD430A2, Thorlabs) with a 250 mm focal length plano-convex lens. Additionally, a long pass filter (937 nm cut off) was placed directly in front of the detector to mitigate any NV center photoluminescence from being detected. The $\frac{1}{2}-$VCSEL chip was pumped with 808~nm light by a fiber coupled laser diode, and the chip's gain peaked around 1040~nm. The BRF was used to tune the lasing mode to 1042.7~nm. Single-frequency operation was verified by a 10 GHz scanning Fabry-Perot interferometer (SA210-8B, Thorlabs). Lastly, a loop antenna was placed inside the VECSEL as close as possible to the surface of the diamond sample, with the 1042~nm cavity mode and 532~nm pump light passing through the center of the loop. The laser is depicted in Fig.~\ref{schematic}. 
\\
\begin{figure}[!t]
 \centering
 \includegraphics[width=\columnwidth]{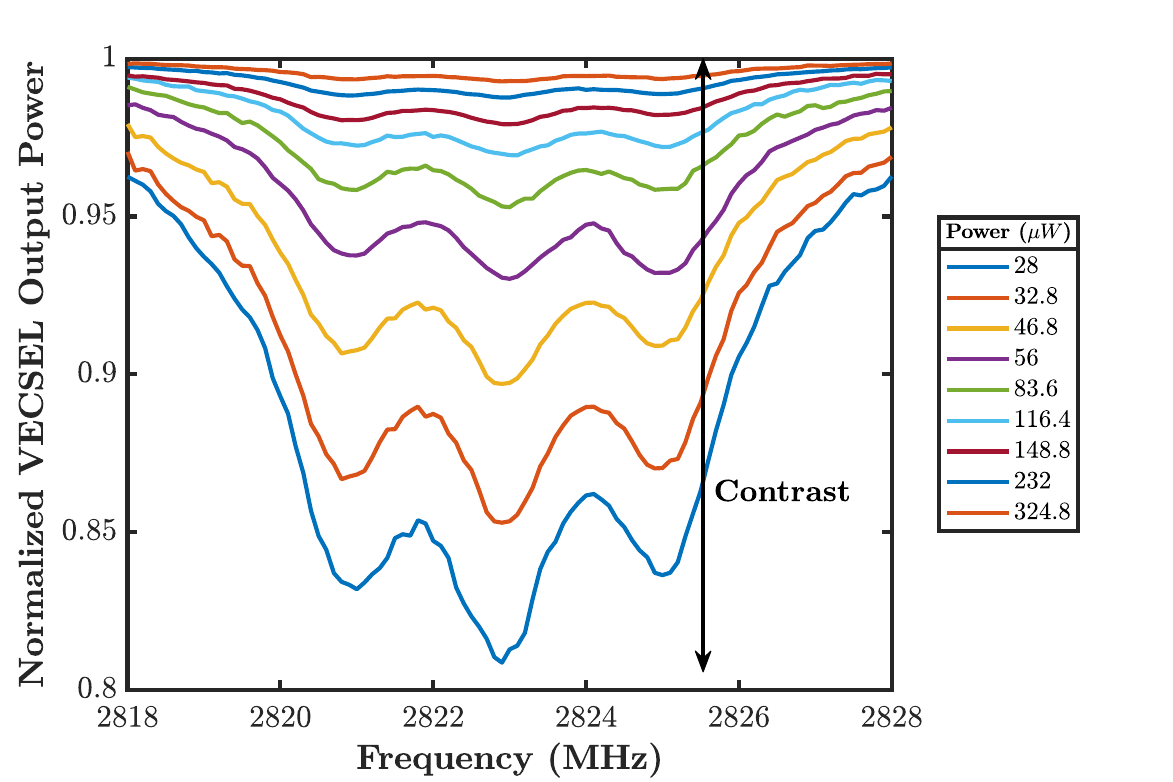}
 \caption{ODMR spectrum of the $\Delta m_s=-1$ transition at 2.8229 GHz, recorded by measuring the output power of the VECSEL. This transition was split from the other orientations using two permanent ring magnets.}
 \label{ODMR Spectrum}
\end{figure}
Two identical permanent ring magnets were placed near the diamond to lift the degeneracy of one of the NV center orientations. To perform this calibration, the diamond was illuminated with 1.2 W of 532~nm pump light. An amplitude modulated 5 kHz square wave microwave probe from a Stanford Research Systems SG 386 signal generator amplified by a Mini-Circuits ZHL-5W-63-S+ amplifier ($\sim$25 dBm delivered to the antenna) was then swept across the relevant frequency range. The resulting voltage from our photodiode was then analyzed by a Stanford Research Systems SR865A lock-in amplifier (LIA) with a 30~ms time constant and 24~dB/oct filter slope. In Fig.~\ref{ODMR Spectrum} the transition corresponding to $\Delta m_s=-1$ is shown. Within the transition, three distinct resonances are observed, corresponding to the hyperfine interaction of the electron and $^{14}N$ nucleus. Based on the transition frequency (2.8229 GHz), we approximate the biasing magnetic field provided by the ring magnets to be 1.682 mT along the NV center axis. Additionally, in Fig. \ref{ODMR Spectrum} the ODMR signal at several laser output powers are shown. The laser output power was measured by a Keithley DMM6500 digital multimeter (DMM) connected in parallel with the LIA. A PID controller (SIM960, Stanford Research Systems) was also connected in parallel with a $\sim$10 Hz RC filter, which adjusted the 808 nm pump power to fix the VECSEL output power about a set point. Approaching lasing threshold, there is a dramatic improvement in the signal contrast, while the resonance linewidth remains unchanged. The resonance linewidths were measured by fitting these data to the sum of three Lorentzian functions, in which the resonant frequencies are spaced by the axial hyperfine constant (2.158 MHz). The resonance linewidth was found to be 1.98 MHz $\pm$ 50.8 kHz with no dependence on output power.
\\
\begin{figure}[!t] 
 \centering
 \includegraphics[width=\columnwidth]{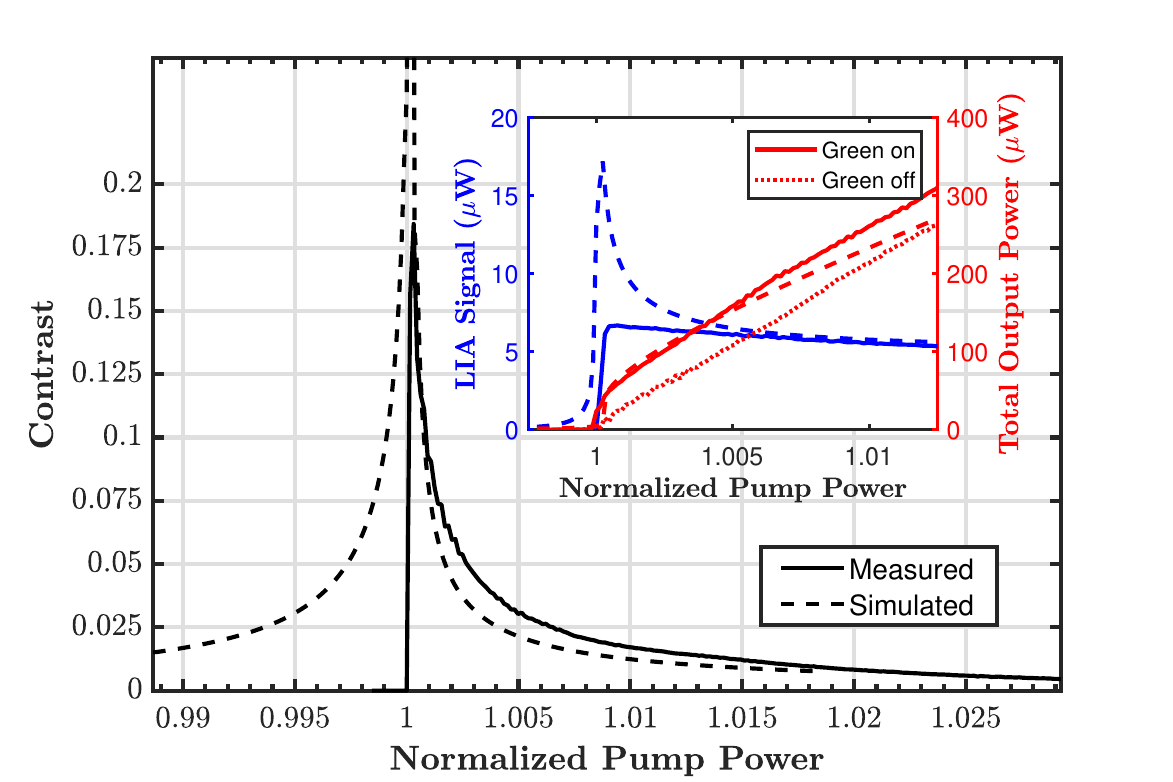}
 \caption{Contrast as a function of normalized 808~nm pump power when probing the transition at 2.8229 GHz. The inset plot shows the ODMR signal (recorded by LIA) and total output power (recorded by DMM). For reference, the total ouput power when the 532~nm pump was off is also included in the inset plot. Additionally, the behavior was reproduced in simulation by including a small saturable absorption term. The simulated behavior is shown as dashed lines.}
 \label{Contrast}
\end{figure}
To approximate the projected PSNL sensitivity and to understand the behavior of the present device with 808~nm pump power, the contrast was measured by applying the same 532 nm pump power and the same amplitude modulated microwave probe on resonance for the transition at 2.8229 GHz. The voltage produced by the photodiode was measured first by the DMM, immediately followed by the LIA with the same time constant and filter slope. Because the ODMR signal read by the LIA is an RMS value of a square wave modulated signal and the voltage measured by the DMM is the average DC value, it can be shown that the contrast is the ratio of the two measured values reduced by a factor of $\sqrt{2}$ (verified by analyzing the photodiode voltage on an oscilloscope and comparing to LIA readings). This is reported in Fig.~\ref{Contrast} as a function of the normalized 808~nm pump power, where $P_{Thresh}$= 4.33 W. Additionally, the ODMR signal and the laser output power are reported in the inset plot. Though the contrast does improve close to threshold, the behavior of the VECSEL near threshold diverges from that predicted by establish theory of infrared LTM.\cite{dumeige_infrared_2019} The LIA signal was observed to decrease farther from threshold, which was a direct result of the nonlinear slope near threshold (Fig.~\ref{Contrast} - inset). This nonlinearity behaves like a saturable absorber, which increases the signal near threshold and returns to a linear relationship at a few percent above threshold. We simulated this behavior with an additional small saturable absorption term (c.f. Ref.~\citenum{Hercher:67}, where $\exp\left(\alpha_0 L_{RT}\right)\sim$10$^{-4}$) modifying the VECSEL photon lifetime\cite{10.1063/1.3222899} which results in a nonlinear power above threshold. We do not attribute this saturable absorption effect to a particular element in the cavity, but effects like this could result from non-uniform pumping of the multi-quantum well gain chip. It should also be noted that this effect was significantly more pronounced when the NV centers were optically pumped. The simulated behavior diverges from our measurement below threshold due to the inclusion of spontaneous emission from the $\frac{1}{2}-$VCSEL gain chip into the lasing mode. This results in an exponential decay of the contrast below threshold; however, the photocurrent generated by the miniscule spontaneous emission reaching the detector was likely below the noise floor of our photodiode. 
\begin{figure}[!t]
 \centering
 \includegraphics[width=\columnwidth]{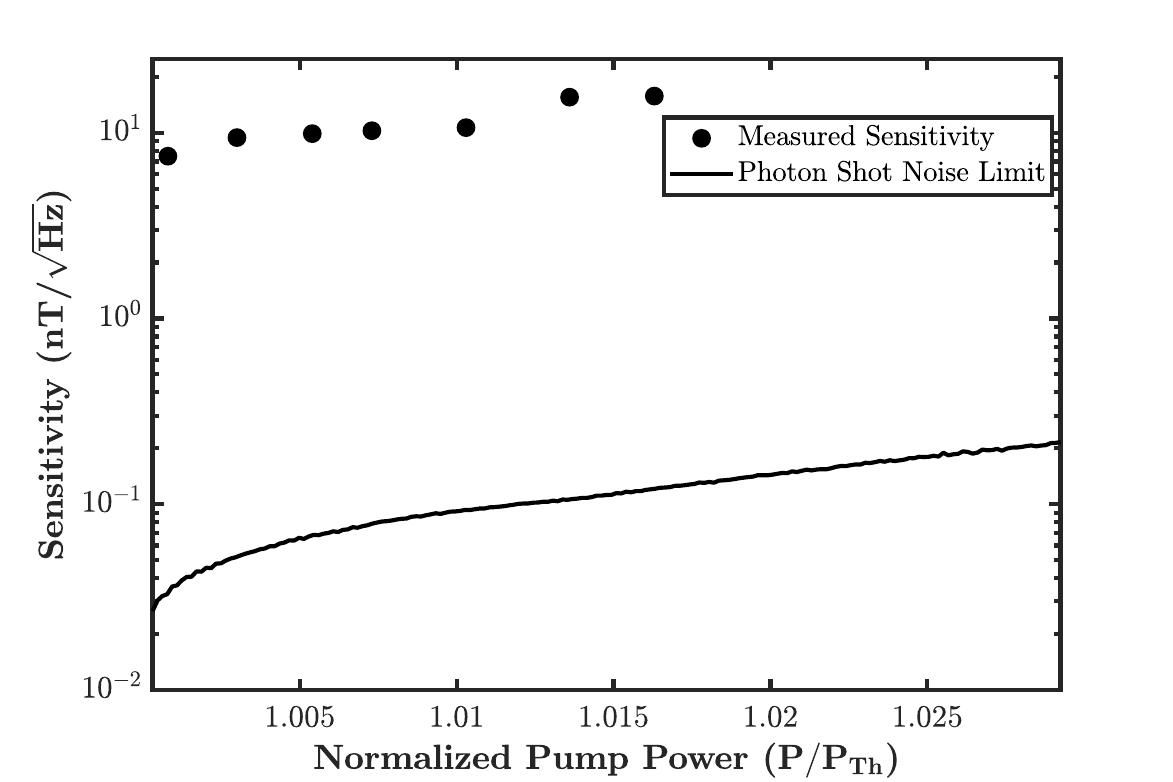}
 \caption{The magnetic sensitivity and the projected photon shot noise limited sensitivity as a function of total detected power at the output of the VECSEL.}
 \label{Sensitivity}
\end{figure}
\\
The projected PSNL sensitivity may be related to the microwave resonance linewidth $\Delta \nu$, ODMR signal contrast $C$, and collected power $P$ by equation \eqref{PSNLeq} (c.f. Ref.~\citenum{barry_sensitivity_2020}).
\begin{equation}
\eta_{PSN}=\frac{4}{3\sqrt{3}}\frac{h}{g_e\mu_B}\frac{\Delta \nu}{C}\sqrt{\frac{E_p}{P}}
 \label{PSNLeq}
\end{equation}
Where $h$ is Planck's constant, $g_e$ is the NV electronic g factor, $\mu_B$ is the Bohr magneton, and $E_p$ is the photon energy. The projected PSNL sensitivity of the device was calculated using the data presented in Fig. \ref{Contrast}, where the microwave resonance linewidth ($\Delta \nu \approx$ 1.98 MHz) is taken to be independent of 808 nm pump power. The resulting projected PSNL sensitivity is shown in Fig. \ref{Sensitivity}. This sensitivity limit showed a dramatic improvement when operating near the lasing threshold. The best result, 26.6~pT/$\sqrt{\textit{Hz}}$, was achieved as close to the lasing threshold as possible. These data show a linear relationship between the projected PSNL sensitivity and the 808 nm pump power of the VECSEL over 1\% above threshold, as predicted by Ref. \citenum{dumeige_infrared_2019}. The saturable absorption behavior introduces a nonlinear enhancement of the projected PSNL sensitivity closer to threshold. 
\begin{figure}[!t]
 \centering
 \includegraphics[width=\linewidth]{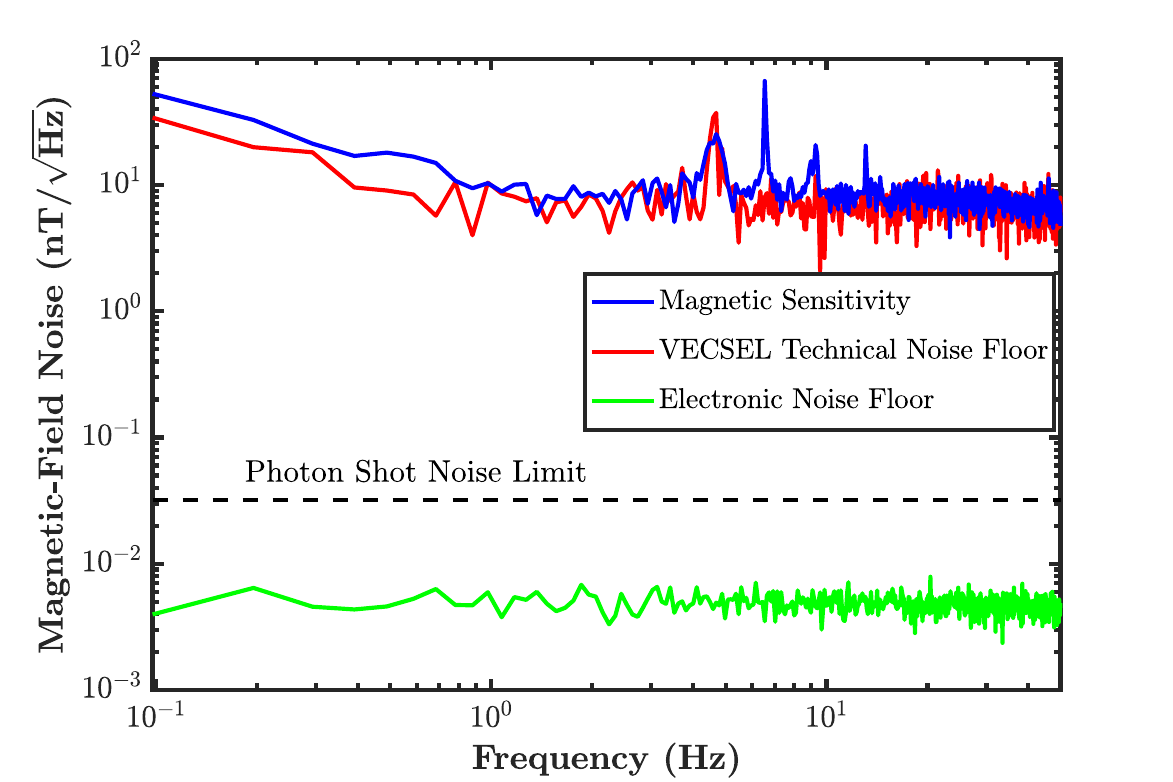}
 \caption{Magnetic-field noise spectrum with 28 $\mu$W of output power. The magnetic sensitivity is seen to be limited by the VECSEL technical noise floor.}
 \label{PSD}
\end{figure}
\\
To approximate the present device's magnetic sensitivity, the same amplitude modulated microwave probe was set to 2.8201 GHz, which lies near the inflection point of the $\Delta m_s=-1$ transition. Additionally, the 532 nm pump power remained the same, but the LIA time constant was reduced to 1 ms to increase the equivalent noise bandwidth to $\sim$78 Hz. The PID feedback loop was used to maintain the VECSEL output power at the same levels as in Fig. \ref{ODMR Spectrum}. Ten 10-second time trace were taken to measure the LIA voltage noise. These time traces were converted to voltage noise spectra, then averaged. The voltage-frequency conversion factor was taken from the slope of the Lorentzian fit of the data in Fig. \ref{ODMR Spectrum} at 2.8201 GHz, and was used in combination with frequency-magnetic-field conversion factor, $\frac{4}{3\sqrt{3}}\frac{h}{g_e\mu_B}$, to convert the LIA voltage noise spectrum to the magnetic-field noise spectrum. Additionally, sets of 10 second time traces were recorded with the microwave probe and green pump off to characterize the VECSEL's technical noise, and with only the detection electronics on to characterize the electronic noise floor. These traces were converted to effective magnetic-field noise spectra to estimate the various noise floors limiting the device. The magnetic-field noise spectrum while the VECSEL was operating at 28 $\mu$W output power is shown in Fig. \ref{PSD}. Additionally, the VECSEL's technical noise floor, the electronic noise floor, and the photon shot noise limit are reported in this figure. The magnetic-field noise spectrum and the VECSEL's technical noise floor are nearly the same, which demonstrates that the present device is heavily limited by the stability of the VECSEL. Spikes in the magnetic-field noise spectrum which are not present in the VECSEL's technical noise spectrum are likely due to ambient magnetic field noise. In the results reported here, we approximate the magnetic sensitivity from 10 Hz - 50 Hz as the mean value of the magnetic-field noise spectrum over this range. The magnetic sensitivity for several pump powers is included in Fig.~\ref{Sensitivity}, and the best magnetic sensitivity was 7.5~nT/$\sqrt{\textit{Hz}}$. The sensitivity did improve by operating close to threshold, however not to the same degree that the projected PSNL sensitivity did. Likely, this is the case because the modulation of the output power due to loss induced by both the NV centers and technical noise sources are enhanced.
\\
In summary, we have demonstrated infrared laser threshold magnetometry by integrating a single crystal diamond plate doped with nitrogen-vacancy centers into a vertical external cavity surface emitting laser. Near threshold operation allowed for significant improvement in the contrast and the projected photon shot noise limited sensitivity, achieving 18.4\% and 26.6~pT/$\sqrt{\textit{Hz}}$, respectively. These results are comparable to other examples of cavity enhance infrared absorption magnetometry\cite{jensen_cavity-enhanced_2014,chatzidrosos_miniature_2017} and cavity enhanced stimulated emission magnetometry\cite{hahl_magnetic-field-dependent_2022}; however, a significantly lower nitrogen-vacancy density was used in the present device than in these works. A further enhancement near threshold of the contrast and photon shot noise limited sensitivity not predicted in previous works was observed due to a saturable absorption behavior of the VECSEL. Further study is needed to diagnose the cause of the effect and how to best take advantage of it. The technical noise of the VECSEL limited the magnetic sensitivity to 7.5 nT/$\sqrt{\textit{Hz}}$. The magnetic sensitivity did not significantly improve when operating near threshold, likely because technical noise sources were amplified. Optimization of output coupling may yield improvements in both the projected photon shot noise limited sensitivity and the magnetic sensitivity. Further improvement of the magnetic sensitivity may be achieved through cavity stabilization and noise mitigation techniques such as, higher frequency power stabilization loops, Pound-Drever-Hall cavity locking, or simultaneous measurement of the technical noise sources and electronic cancellation. More complex probing techniques such as three tone FM probes \cite{barry_optical_2016}, dual resonance probes \cite{fescenko_diamond_2020}, pulsed ODMR, or Ramsey magnetometry \cite{barry_sensitivity_2020} may yield a further enhancement. Higher nitrogen-vacancy densities or increased sample thicknesses may improve the contrast, projected photon shot noise limited sensitivity, and magnetic sensitivity. Lastly, engineering the gain of the $\frac{1}{2}-$VCSEL chip to provide a higher slope efficiency may also improve the performance of the device.\\
The authors would like to acknowledge the support of the Air Force Research Laboratory (UES contract no. S-119-005-002) and the National Science Foundation (intern funding support).
\section*{References}
%

\end{document}